\documentclass[aps,pre,twocolumn,groupedaddress,floatfix]{revtex4-1}
\usepackage{color}
\usepackage{tabularx}
\usepackage{epsfig}
\usepackage{amsmath}
\usepackage{graphicx}

\begin{document}

\title{Ising Spin Glasses in dimension five}

\author{P. H.~Lundow} \affiliation{Department of Mathematics and
  Mathematical Statistics, Ume{\aa} University, SE-901 87, Sweden}

\author{I. A.~Campbell} \affiliation{Laboratoire Charles Coulomb
  (L2C), UMR 5221 CNRS-Universit\'e de Montpellier, Montpellier,
  F-France.}

\date{\today}

\begin{abstract}
Ising spin glass models with bimodal, Gaussian, uniform and Laplacian
interaction distributions in dimension five are studied through
detailed numerical simulations. The data are analyzed in both the
finite-size scaling regime and the thermodynamic limit regime. It is
shown that the values of critical exponents and of dimensionless
observables at criticality are model dependent. Models in a single
universality class have identical values for each of these critical
parameters, so Ising spin glass models in dimension five with
different interaction distributions each lie in different universality
classes.  This result confirms conclusions drawn from measurements in
dimension four and dimension two.
\end{abstract}


\maketitle

\section{Introduction}\label{sec:I}
The statistical physics of second order transitions has been
intensively studied in standard systems exemplified by pure
ferromagents, and a thorough understanding of the critical behavior has
been reached based on Renormalization Group Theory (RGT).  RGT
provides an elegant explanation of the universality of critical
exponents, which is the property that all systems within the same
universality class (determined only by the physical dimension $d$ and
the spin dimension $N$) have identical values for each critical
exponent and for characteristic dimensionless critical parameters. It
has been implicitly or explicitly assumed that in spin glasses the
form of the interaction distribution is not a relevant parameter for
the determination of the universality class, so that in particular all
Ising Spin Glasses (ISGs) in a given dimension are expected to have
the same critical exponents and critical parameters.  The ISG
situation is in fact much less clear cut; it has been stated that a
fundamentally different theoretical approach to transitions is
required \cite{parisi:01,castellana:11,angelini:13}.  We have found
from numerical studies on Ising spin glasses (ISGs) in dimensions $4$
and $2$ having bimodal and Gaussian interaction distributions
\cite{lundow:15,lundow:15a,lundow:16} that the critical exponents and
the critical values for dimensionless constants are not identical for
the two models in a given dimension but that they vary with the
interaction distribution. It was concluded that the universality class
of an ISG depends not only on the physical dimension of the system but
also on the form of the interaction distribution.

Here numerical simulation data on ISGs in dimension $5$ are presented
and analysed. We are aware of no analogous simulation measurements on
ISGs in dimension $5$, but some of the present measurements can be
compared to results on the same models obtained from the High
Temperature Series Expansion (HTSE) technique
\cite{klein:91,daboul:04}. Again, as in dimensions $4$ and $2$ the
values for critical dimensionless constants and for the critical
exponents are found to vary with the form of the interaction
distribution, confirming that the non-universality conclusion reached
for ISGs can be generalized.

Dimension $5$ is close to the ISG upper critical dimension $d=6$. For
reference, the $\epsilon$-expansion ISG exponent values to leading
order in $\epsilon = 6-d$ \cite{gardner:84} are $\gamma =1+(6-d)$,
$\nu = 1/2+5(6-d)/12$ and $\eta = -(6-d)/3$, so for dimension $5$ the
leading order exponent values are $\gamma = 2$, $\nu = 11/12 \approx
0.92$, and $\eta = -1/3 \approx -0.33$. The terms of higher order in
$\epsilon$ are strong and no summations over all terms are known.
There are no interaction distribution dependent terms in the standard
$\epsilon$-expansion expressions. The leading order
$\epsilon$-expansion exponent values are all in rough agreement with
but are about $25\%$ stronger than the range of numerical estimates
for the 5D ISG exponents given below in the Conclusion, Table I,
where for instance the $\gamma$ estimates run from $1.73(2)$ for the
bimodal interaction model to $1.49(2)$ for the Laplacian interaction
model.

\section{Historical note}\label{sec:II}
In $1894$ Van der Waals introduced the concept of critical exponents,
in the context of transitions in liquids; he derived values for the
exponents in terms of what is now called a mean field theory
\cite{vanderwaals:94}.  His student Verschaffelt made very precise
experimental measurements on capillarity, and in $1900$ published
experimental estimates for the exponents which were not equal to the
mean field values \cite{verschaffelt:00}. His results were ignored for
sixty years because they had no theoretical support (see
Ref.~\cite{levelt:76} for an excellent historical account). The
situation only changed with Onsager's analytic proof of non-mean field
exponent values in the $2$D Ising model \cite{onsager:44}, which
finally led on to the establishment of the principle of universality,
within the RGT concept \cite{wilson:71}.

Verschaffelt employed temperature dependent effective exponents in his
analyses. Effective exponent analyses were re-introduced much later
for experimental \cite{kouvel:64} and numerical
\cite{orkoulas:00,butera:02} ferromagnetic data. Below we also will
use effective exponents in the analysis of ISG simulation data. We
obtain results which are firmly established empirically but for which
a theoretical explanation is for the moment lacking.

\section{Simulation measurements}\label{sec:III}
We are aware of no previous publications of precise simulation data on
ISGs in dimension $5$.  The standard ISG Hamiltonian is
\begin{equation}
  \mathcal{H}= - \sum_{ij}J_{ij}S_{i} S_{j}
  \label{ham}
\end{equation}
with the near neighbor symmetric distributions normalized to $\langle
J_{ij}^2\rangle=1$. The normalized inverse temperature is $\beta =
(\langle J_{ij}^2\rangle/T^2)^{1/2}$. The Ising spins live on simple
hyper-cubic lattices with periodic boundary conditions. The spin
overlap parameter is defined as usual by
\begin{equation}
  q =\frac{1}{L^{d}}\sum_{i} S^{A}_{i} S^{B}_{i}
\end{equation}
where A and B indicate two copies of the same system. We have studied
the symmetric bimodal ($\pm J$), Gaussian, uniform
($P(J)=1/[2\cdot 3^{1/2}]$ for $-3^{1/2} < J < 3^{1/2}$) and Laplacian
$P(J) = 2^{1/2}\exp(-2^{1/2}|J|)$ distribution ISG models in dimension
$5$.

The simulations were carried out using the exchange Monte-Carlo method
for equilibration using so called multi-spin coding, on $2^{12}$
individual samples at each size from $L=3$ to $L=10$ for the bimodal
and Gaussian models. (It can be noted that an $L=10$ sample in $d=5$
contains more spins than an $L=46$ sample in $d=3$, so the simulations
are laborious. However, see the Thermodynamic Limit (ThL) figure of
merit discussion in Section~\ref{sec:VI}. For the uniform distribution
model $2^{12}$ samples were studied up to $L=6$, and $2^9$ samples up
to $L=9$, and for the Laplacian model $2^{12}$ samples were studied up
to $L=6$ and $2^{9}$ samples up to $L=8$.  An exchange was attempted
after every sweep with a success rate of at least $30\%$. At least
$40$ temperatures were used forming a geometric progression reaching
up to $\beta_{\max}=0.42$ in the bimodal model, $\beta_{\max}=0.45$ in
the Gaussian model, $\beta_{\max}=0.45$ in the uniform model and
$\beta_{\max}=0.45$ in the Laplacian model.  This ensures that our
data span the critical temperature region which is essential for the
finite-size scaling (FSS) analyses. Near the critical temperature the
$\beta$ step length was at most $0.003$. The various systems were
deemed to have reached equilibrium when the sample average
susceptibility for the lowest temperature showed no trend between
runs. For example, for $L=10$ this means about $200000$ sweep-exchange
steps.

After equilibration, at least $200000$ measurements were made for each
sample for all sizes, taking place after every sweep-exchange step.
Data were registered for the energy $E(\beta,L)$, the correlation
length $\xi(\beta,L)$, for the spin overlap moments $\langle |q|
\rangle$, $\langle q^2\rangle$, $\langle |q|^3\rangle$, $\langle
q^4\rangle$ and the corresponding link overlap $q_{\ell}$ moments.  In
addition the correlations $\langle E(\beta,L),U(\beta,L)\rangle$
between the energy and observables $U(\beta,L)$ were also registered
so that thermodynamic derivatives could be evaluated using the
thermodynamic relation $\partial U(\beta,L)/\partial \beta = \langle
U(\beta,L), E(\beta,L)\rangle-\langle U(\beta,L) \rangle\langle
E(\beta,L)\rangle$ where $E(\beta,L)$ is the energy
\cite{ferrenberg:91}.  Bootstrap analyses of the errors in the
derivatives as well as in the observables $U(\beta,L)$ themselves were
carried out.
We follow the same analysis strategy for the 5D ISGs as for the $4$D
ISGs \cite{lundow:14,lundow:15a}.

\section{Finite size scaling}\label{sec:IV}
The usual approach to critical parameter measurements through
simulations is to study the size dependence of dimensionless
observables $Q(\beta,L)$ (generally the Binder cumulant $g(\beta,L) =
(3- \langle q^4\rangle/\langle q^2\rangle^2)/2$ and the normalized
correlation length $\xi(\beta,L)/L$) in the regime very near the
critical point. $g(\beta,L)$ must saturate at $g(\beta,L)=1$ for
$\beta \gg \beta_{c}$ which is not the case for $\xi(\beta,L)/L$. It
can be noted that we find critical $g(\beta_{c})$ values much lower in
5D ISGs than in $3$D or even in $4$D ISGs, so the 5D $g(\beta,L)$
data have space to "fan out" beyond $\beta_{c}$ making this parameter
more efficient for critical regime analyses in 5D than in the other
dimensions.  The typical FSS expression, valid in the near critical
region if there is a single dominant scaling correction term, is :
\begin{equation}
  Q(\beta,L) = Q_{c}+ AL^{-\omega} +B(\beta-\beta_{c})L^{1/\nu}
\end{equation}
where $\nu$ is the correlation length critical exponent and $\omega$
is the exponent of the leading finite size correction term. For any
dimensionless parameter $Q$ the $Q_{c}$ critical values are identical
for all systems within a universality class.  From the HTSE and
thermodynamic limit (ThL) data which we will discuss later the 5D
correction exponent is typically $\omega \approx 1.0$ in the different
models.

We will use the finite size scaling measurements as one method to
estimate the critical inverse temperatures $\beta_{c}$, together with
the dimensionless parameter values $Q_{c}$ at criticality extrapolated
to the infinite size limit.  The critical exponent $\nu$ can be
estimated from the derivatives at criticality through
\begin{equation}
  \frac{\partial Q(\beta,L)}{\partial\beta}\Big\vert_{\beta_c} =
  A_{Q}L^{1/\nu}\left(1 +a_{Q}L^{-\omega} + \cdots\right)
  \label{dQdb}
\end{equation}

The critical exponent $\eta$ can be estimated through
\begin{equation}
  \frac{\chi(\beta_{c},L)}{L^{2}} = A_{\chi}L^{-\eta}\left(1 +a_{\chi}L^{-\omega} +
    \cdots\right)
  \label{sovLsq}
\end{equation}

For the present analysis we have recorded the FSS behavior of various
dimensionless parameters in addition to the Binder cumulant
$g(\beta,L)$ and the correlation length ratio $\xi(\beta,L)/L$.  The
dimensionless parameter $W(\beta,L)$ for Ising ferromagnets was
introduced in Ref.~\cite{lundow:10}.  In the ISG context the parameter
$W_{q}(\beta,L)$ is defined by
\begin{equation}
  W_{q}(\beta,L) = \frac{1}{\pi-2} \left(\frac{\pi\,\lbrack\langle
    |q|\rangle\rbrack^2}{\lbrack\langle q^2\rangle\rbrack} - 2\right)
  \label{Wqdef}
\end{equation}

In the same spirit we will also make use of other dimensionless
parameters
\begin{equation}
  h(\beta,L) = \frac{1}{\sqrt{\pi}-\sqrt{8}}
  \left(\sqrt{\pi}\,\frac{\lbrack\langle
    |q^3|\rangle\rbrack}{\lbrack\langle q^2\rangle\rbrack^{3/2}} -
  \sqrt{8}\right)
  \label{W32}
\end{equation}

\begin{equation}
  P_{W} = \left\lbrack\frac{\langle|q|\rangle^{2}}{\langle
    q^{2}\rangle}\right\rbrack
  \label{PW}
\end{equation}
and the skewness 
\begin{equation}
  P_{\mathrm{skew}} = \left\lbrack\frac{\langle |q|^3\rangle}{\langle
    q^2\rangle^{3/2}}\right\rbrack
  \label{Pskew}
\end{equation}
which also have analogous scaling properties.

\section{Thermodynamic derivative peak analysis}\label{sec:V}
The thermodynamic derivative peak analysis can also be an efficient
method for analyzing data in a ferromagnet or an ISG.
Near criticality in a ferromagnet, for a number of standard
observables $Q$ the heights of the peaks of the thermodynamic
derivatives $\partial Q(\beta,L)/\partial \beta$ scale for large $L$
as \cite{ferrenberg:91,weigel:09}
\begin{equation}
  D_{\max}(L) = \frac{\partial Q(\beta,L)}{\partial
  \beta}\Big\vert_{\max} \propto L^{1/\nu}\left(1+
  aL^{-\omega/\nu}\right)
  \label{dUdbmax}
\end{equation}
The observables used for $Q(\beta,L)$ can be for instance the Binder
cumulant $g(\beta,L)$ or the logarithm of the finite size
susceptibility $\ln(\chi(\beta,L))$ \cite{ferrenberg:91}.  Without
needing a value of $\beta_{c}$ as input, the large $L$ peak height
$D_{\max}(L)$ against $L$ plot provides $1/\nu$ directly, to within
scaling corrections.

In addition, the temperature location of the derivative peak
$\beta_{\max}(L)$ scales as
\begin{equation}
  \beta_{c}-\beta_{\max}(L) \propto
  L^{-1/\nu}\left(1+bL^{-\omega/\nu}\right)
\end{equation}

We note that the inverse of the derivative peak height $1/D_{\max}(L)$
and the peak location temperature difference
$[\beta_{c}-\beta_{\max}(L)]$ are both proportional to
$L^{-1/\nu}(1+aL^{-\theta/\nu} +\cdots)$ (with the leading correction
terms having different pre-factors). Then $\beta_{\max}(L)$ plotted
against $1/D_{\max}(L)$ must tend linearly towards the intercept
$\beta_c$ as $1/D_{\max}(L)$ tends to zero for large $L$. All plots of
the same type for different observables $Q$ should extrapolate
consistently to the true $\beta_c$. The leading correction is
eliminated to first order and together with the higher order
corrections only appears as a modification to the straight line for
small $L$. Provided that the peaks for the chosen observable fall
reasonably close to $\beta_{c}$ these data can be much simpler to
analyse than those from the crossing technique. For ferromagnets,
Ferrenberg and Landau \cite{ferrenberg:91} found this form of analysis
significantly more accurate than the standard Binder cumulant crossing
approach.

In the ISG context exactly the same methodology can be used as in the
ferromagnet \cite{lundow:15}. Because the exponent $\nu$ is relatively
small in 5D ISGs this technique is an efficient independent method
for estimating $\beta_c$.  As far as we are aware this analysis has
not been used previously by other authors in ISGs.

\section{Thermodynamic limit scaling}\label{sec:VI}
The high temperature series expansion for the spin glass
susceptibility of an ISG with a symmetrical interaction distribution
can be written \cite{daboul:04}
\begin{equation}
  \chi(\beta^2) = 1 + a_{1}\beta^2 + a_{2}\beta^4 + \cdots
\end{equation}
Only even terms in powers of $\beta$ exist because of the symmetry
between positive and negative interactions so $\beta^2$ rather than
$\beta$ is the natural thermal scaling variable
\cite{singh:86,klein:91,daboul:04,campbell:06}. (An equivalent natural
scaling variable which has been generally used for HTSE analyses on
ISGs with symmetric bimodal interaction distributions is
\cite{singh:86,klein:91} $w = 1
-\tanh(\beta)^2/\tanh(\beta_{c})^2$. The discussion above holds
throughout with $w$ replacing $\tau$. The exponents of course remain
the same though the factors $C, a$ etc. are modified.)  In principle
an infinite set of exact HTSE factors $a_{n}$ exist.  In practice
terms in different ISG models have been calculated at best up to
$n=15$ (see Refs.~\cite{singh:86,klein:91,daboul:04}).  Then according
to Darboux's first theorem \cite{darboux:78} the asymptotic form of
the sum of the entire series (all terms to infinite $n$) is eventually
dominated by the closest singularity to the origin, which in the
simplest case is the physical singularity, so near $\beta_{c}^2$
\begin{equation}
  \chi(\beta^2) = C_{\chi}\left[1-(\beta/\beta_{c})^2\right]^{-\gamma}
\end{equation}
with $\beta_{c}^2$ being the inverse critical temperature squared and
$\gamma$ the standard critical exponent.

It is thus natural to adopt $\tau = 1-(\beta/\beta_{c})^2$ as the
scaling variable in analyses of ThL ISG simulation data just as in the
HTSE analyses\cite{daboul:04,campbell:06}.  Then the Wegner scaling
expression \cite{wegner:72} for the ThL ISG susceptibility is
\begin{equation}
  \chi(\tau) = C_{\chi}\tau^{-\gamma}\left(1+a_{\chi}\tau^{\theta} +
    b_{\chi}\tau^\theta{'} + \cdots\right)
  \label{wegchi}
\end{equation}
where $\theta=\nu\omega$ is the leading thermal correction exponent
and the second term is generally analytic.  The standard RGT scaling
variable $t = (T-T_{c})/T_{c}$ is often used for ISG simulation
analyses close to criticality, but this scaling variable is not
convenient at higher temperatures as $t$ diverges at infinite
temperature while $\tau$ tends to $1$. Also the temperature appears as
$T$ not $T^2$ so $t$ is only appropriate for ISGs as an approximation
near $\beta_{c}$.

The HTSE second moment of the ISG spin-spin correlations $\mu_{2} =
\sum r^2\langle S_{0}.S{r}\rangle$ is of the form (see
Ref.~\cite{butera:02,campbell:08} for the ferromagnetic case)
\begin{equation}
  \mu_{2}(\beta^2) = \beta^2\left(z + b_{1}\beta^2 + b_{2}\beta^4 + \cdots\right)
\end{equation}
where $z$ is the number of near neighbors.  The ThL $\mu_{2}$ diverges
at $\beta_{c}$ as $\tau^{-(\gamma+2\nu)}$. Then, invoking again
Darboux's theorem to link the series within the brackets to the
critical divergence, the appropriate scaling form can be written as
\begin{equation}
  \mu_{2}(\beta^2) =
  C_{\mu}z\beta^{2}\tau^{-(\gamma+2\nu)}\left(1+a_{\mu}\tau^{\theta}
    +\cdots\right)
\end{equation}
As the ThL second moment correlation length is defined through
$\mu_{2} = z\chi(\beta)\xi(\beta)^{2}$, the Wegner form for the
normalized ISG ThL correlation length can be written
\cite{campbell:06}
\begin{equation}
  \xi(\beta)/\beta = C_{\xi}\tau^{-\nu}\left(1 + a_{\xi}\tau^{\theta} +
    b_{\xi}\tau + \cdots\right)
\end{equation}
It is important to note the factor $\beta$ which normalizes
$\xi(\beta)$ in this equation.

The form of susceptibility scaling outlined here for ISGs was used
from the earliest HTSE studies of critical behavior in ferromagnets
and then in ISGs
Refs.~\cite{fisher:67,butera:02,klein:91,daboul:04}. The analogous
normalized correlation length form was introduced explicitly in
Ref.~\cite{campbell:06}.

The full HTSE sum is by construction in the (infinite $L$)
Thermodynamic limit (ThL) but extrapolations from high temperature
must be made in order to estimate behavior at criticality, because the
complete series is not available \cite{daboul:04}.  Simulation data
are necessarily taken at finite $L$, but can be considered as also
being in the ThL as long as $L \gg \xi(\beta)$ where $\xi(\beta)$ is
the ThL correlation length. The ThL envelope curves can generally be
recognized by inspection of the data plots. As a rule of thumb, the
condition $L > 6\xi(\beta)$ can generally be taken as sufficient, with
observables independent of $L$ and equal to the ThL values as long as
this condition is satisfied. The simulation data supplement and extend
the HTSE data. As $\xi(\beta) \sim \beta[1-(\beta/\beta_c)^2]^{-\nu}$
in ISGs the ThL condition can be written approximately in terms of a
figure of merit; if $\tau_{\min}$ is the lowest reduced temperature to
which the ThL condition holds for size $L$,
\begin{equation}
  \tau_{\min} \approx (L/6\beta_{c})^{-1/\nu}
  \label{taum}
\end{equation}
In dimension $5$ with $\beta_{c} \approx 0.4$ and $\nu \approx 0.7$
the condition implies $\tau_{\min} \approx 0.15$ if the largest size
used is $L = 10$.  This $\tau_{\min}$ corresponds to a temperature
within $8\%$ of the critical temperature. It can be underlined that in
dimension $3$ with the appropriate parameters for ISGs, $\beta_{c}
\approx 1$, $\nu \approx 2.5$, to reach $\tau_{\min} \approx 0.15$
would require sample sizes to $L \approx 300$, far beyond the maximum
sizes which have been studied numerically up to now in 3D ISGs.  The
ISG ThL regime can be studied numerically reasonably close to
criticality in dimension $5$ (and dimension $4$) but the situation is
much more delicate in dimension $3$.

Temperature and size dependent susceptibility and correlation length
effective exponents, valid over the entire paramagnetic regime, can be
defined by
\begin{equation}
  \gamma(\tau,L) = - \partial\ln\chi(\tau,L)/\partial\ln\tau
  \label{gameff}
\end{equation}
and
\begin{equation}
  \nu(\tau,L) = -\partial\ln[\xi(\tau,L)/\beta]/\partial\ln\tau
  \label{nueff}
\end{equation}
The critical limits are $\gamma(0,\infty) =\gamma$ and $\nu(0,\infty)
=\nu$; extrapolations must be made to estimate the critical exponents
from HTSE or simulation data.  In simple hyper-cubic lattices of
dimension $d$ where $z=2d$ the exact ISG high temperature limits for
all $L$ are $\gamma(1,L) = 2d\beta_{c}^2$, and $\nu(1,L) = (d -
K/3)\beta_{c}^2$ where $K$ is the kurtosis of the interaction
distribution ($K=1$ for the bimodal distribution, $K=3$ for the
Gaussian distribution, $K=9/5$ for the uniform distribution, and $K=6$
for the Laplacian distribution).

The value of $\beta_c$ enters implicitly into the definitions of
$\gamma(\tau,L)$ and $\nu(\tau,L)$ in Eq.~\ref{gameff} and
Eq.~\ref{nueff} through the definition of $\tau$, so it is important
to have well established estimates for $\beta_{c}$ for the $\gamma$
and $\nu$ effective exponent analyses.

Turning to the exponent $\eta$, the temperature dependent effective
$\eta(\beta,L)$ can be estimated through
\begin{equation}
  2 - \eta(\beta,L) = \frac{\partial\ln\chi(\beta,L)}{\partial
  \ln[\xi(\beta,L)/\beta]} = \frac{\gamma(\beta,L)}{\nu(\beta,L)}
  \label{etaeff}
\end{equation}

Alternatively one can make a log-log plot of $y(\beta,L) =
\chi(\beta^2,L)/[\xi(\beta^2,L)/\beta]^2$ against $x(\beta,L) =
\xi(\beta^2,L)/\beta$. At high temperatures and for all $L$,
$x(\beta,L)$ and $y(\beta,L)$ both tend to $1$ as $\beta$ tends to
$0$. For large $L$ and temperatures near criticality the slope of the
ThL envelope curve $\partial \ln y(\beta,L)/\partial \ln x(\beta,L)$
tends to the critical exponent $-\eta$ in the limit $\beta \to
\beta_{c}$ where both $y(\beta,L)$ and $x(\beta,L)$ diverge. With an
appropriate fit function, extrapolation of the ThL envelope curve to
the large $L$ limit leads to an estimate for $\eta$ purely from ThL
data, without invoking the FSS estimate for $\beta_{c}$.


\section{Privman-Fisher scaling}\label{sec:VII}
The Privman-Fisher scaling ansatz \cite{privman:84} for an observable
$Q(\beta,L)$ can be written in the simple general form
\begin{equation}
  Q(\beta,L)/Q(\beta,\infty) = F[L/\xi(\beta,\infty)]
  \label{PF}
\end{equation}
where Wegner thermal correction terms are implicitly included in
$Q(\beta,\infty)$ and $\xi(\beta,\infty)$. A leading finite size
correction term can be introduced \cite{calabrese:03} :
\begin{equation}
  \frac{Q(\beta,L)}{Q(\beta,\infty)} =
  F\left[L/\xi(\beta,\infty)\right]\left(1 +
  \frac{G_{Q}\left[L/\xi(\beta,\infty)\right]}{L^{\omega}}\right)
\end{equation}
For given values of the critical inverse temperature and exponents
$\beta_{c}$, $\nu$ and $\eta$, assuming the leading ThL ISG extended
scaling expressions $\chi(\beta,\infty) \propto
[1-(\beta/\beta_{c})^2]^{-\gamma}$ and $\xi(\beta,\infty)\propto
\beta[1-(\beta/\beta_{c})^2]^{-\nu}$ are valid and ignoring Wegner
correction terms, the basic Privman-Fisher ansatz for the
susceptibility can be readily transformed into
\begin{equation}
  \frac{\chi(\beta,L)}{(L/\beta)^{2-\eta}} = \mathcal{F}
       [|(1-(\beta/\beta_{c})^2)|(L/\beta)^{1/\nu}]
  \label{PFextscal}
\end{equation}
as applied in \cite{campbell:06,hukushima:09}.  This extended scaling
form is less sensitive to the precise values of the critical
parameters than is the ThL scaling and does not contain the correction
terms. However it allows one to scale all the data, not only those
from the ThL regime, but also from the crossover regime between the
ThL and FSS regimes, from the critical regime, and even from the
region to temperatures rather below the critical temperature. Below it
will be seen that very acceptable scaling is observed for the data
from each of the four models studied, when the appropriate scaling
parameters are used. This shows that the data for all $L$ and for all
temperatures from infinity down to below $T_c$ can be encapsulated in
the scaling expression \eqref{PFextscal}, adjusting only the three
critical parameters $\beta_{c}$, $\nu$ and $\eta$. If Wegner
correction terms have been estimated from ThL scaling these can be
introduced to improve the scaling but their influence will only be
felt well outside the critical region.

\section{The 5D Gaussian distribution ISG model}\label{sec:VIII}
\begin{figure}
  \includegraphics[width=3.5in]{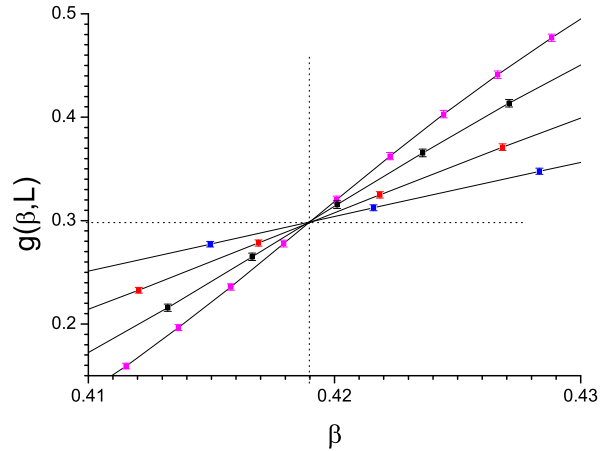}
  \caption{(Color on line) Gaussian 5D ISG model. Even $L$ Binder
    cumulants $g(\beta,L)$ against inverse temperature $\beta$, $L=4$,
    $6$, $8$ and $10$ (top to bottom on the left).}
  \protect\label{fig1}
\end{figure}

\begin{figure}
  \includegraphics[width=3.5in]{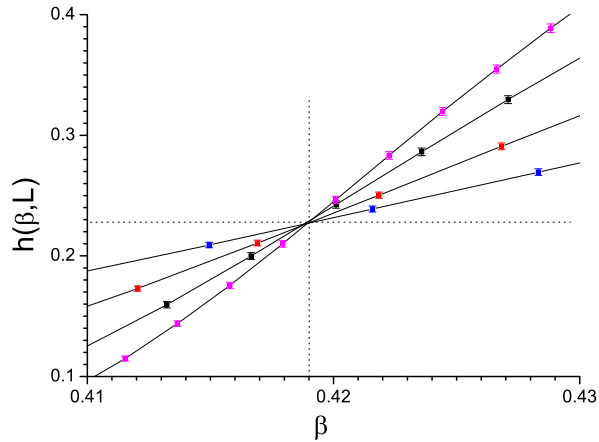}
  \caption{(Color on line) Gaussian 5D ISG model. Even $L$ data for
    the observable $h(\beta,L)$ against $\beta$, $L=4$, $6$, $8$ and
    $10$ (top to bottom on the left).} \protect\label{fig2}
\end{figure}

\begin{figure}
  \includegraphics[width=3.5in]{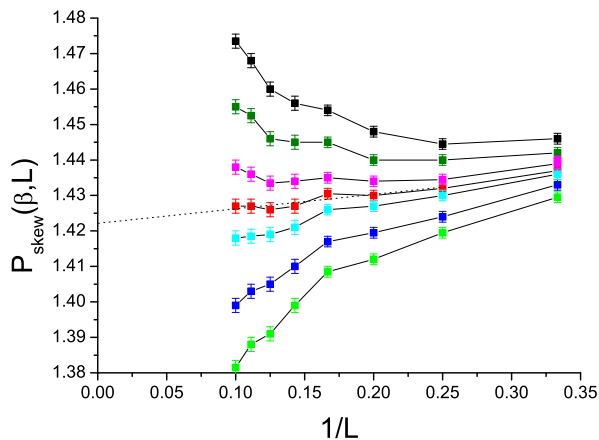}
  \caption{(Color on line) Gaussian 5D ISG. $P_{\mathrm{skew}}(\beta,L)$
    against $1/L$ for fixed $\beta$, $\beta= 0.424$, $0.422$, $0.420$,
    $0.419$, $0.418$, $0.416$ and $0.414$ (top to bottom).  Dashed
    line : estimated criticality.}  \protect\label{fig3}
\end{figure}

\begin{figure}
  \includegraphics[width=3.5in]{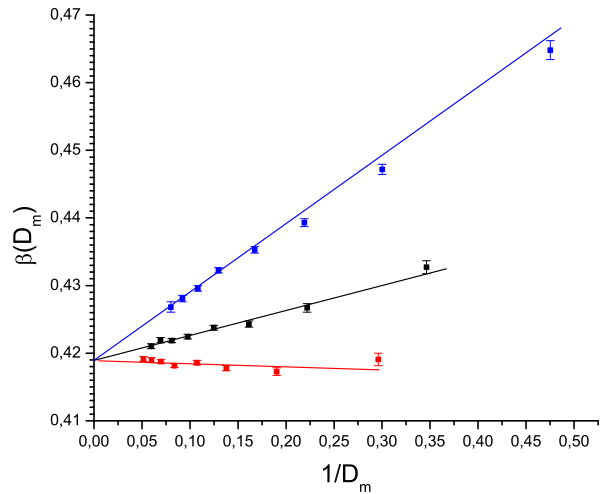}
  \caption{(Color on line) Gaussian 5D ISG. Peak location $y =
    \beta_{m}$ against inverse peak height $x = 1/D_{m}$ for the
    derivatives $\partial P_{W}/\partial \beta$, $\partial
    h/\partial \beta$ and $\partial g/\partial \beta$ (top to
    bottom).  Sizes $L= 3$, $4$, $5$, $6$, $7$, $8$, $9$ and $10$
    (increasing to the left). For each observable the points
    extrapolate to $y(x) = \beta_{c}$ at the lintercept, see text.  }
  \protect\label{fig4}
\end{figure}

For the Gaussian distribution model, the FSS Binder parameter
$g(\beta,L)$ data and the parameter $h(\beta,L)$ both happen to show
no visible correction to scaling at criticality, Figs.~\ref{fig1} and
\ref{fig2}. This provides us with consistent and accurate estimates
$\beta_{c}= 0.4190(3)$, $g_{c}= 0.300(2)$ and $h_{c}=0.225(1)$. The
data for the other dimensionless parameters in the form of fixed
temperature plots show only weak corrections to scaling. They are all
consistent with $\beta{c} = 0.419$ and $\omega \approx 1$. As the
finite size corrections are weak the analyses are rather insensitive
to the assumed value for $\omega$, see for instance
Fig.~\ref{fig3}. The critical value estimates for the dimensionless
parameters are listed in the Conclusion, Table I.
Data for the locations of thermodynamic derivative peaks are shown in
Fig.~\ref{fig4}. They are also all consistent with $\beta_{c} = 0.419$.

\begin{figure}
  \includegraphics[width=3.5in]{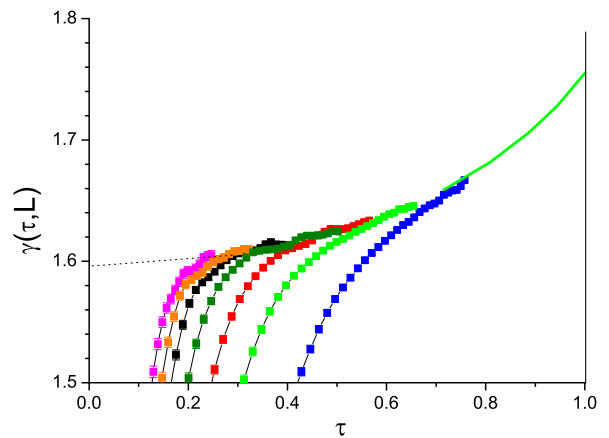}
  \caption{(Color on line) Gaussian 5D ISG. Effective exponent
    $\gamma(\tau,L)$ as function of $\tau$ with $\beta_c =
    0.419$. Points : simulation data for $L=10$, $9$, $8$, $7$, $6$,
    $5$ and $4$ (left to right). Dashed curve: fit. Continuous (green)
    curve on the right : calculated by summing the HTSE tabulation of
    \cite{daboul:04}.}  \protect\label{fig5}
\end{figure}

\begin{figure}
  \includegraphics[width=3.5in]{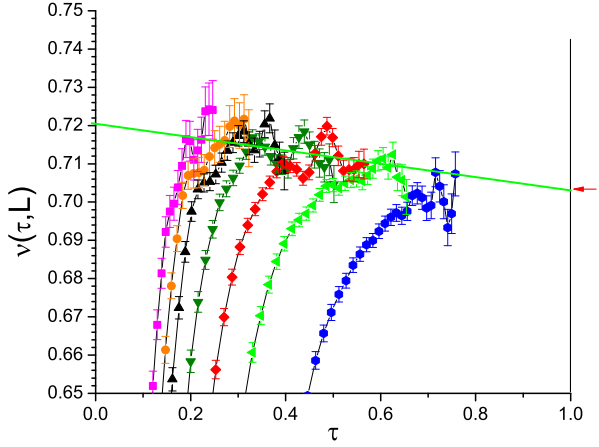}
  \caption{(Color on line) Gaussian 5D ISG. Effective exponent
    $\nu(\tau,L)$ as function of $\tau$ with $\beta_c = 0.419$. Points
    : simulation data for $L=10$, $9$, $8$, $7$, $6$, $5$ and $4$
    (left to right).  Continuous (green) curve : fit.}
  \protect\label{fig6}
\end{figure}

The effective exponents $\gamma(\tau,L)=\partial
\ln\chi(\tau,L)/\partial\ln\tau$ and $\nu(\tau,L)=\partial
\ln[\xi(\tau/L)/\beta]/\partial\ln\tau$ with $\beta_{c}$ fixed at
$0.419$ are shown in Figs.~\ref{fig5} and \ref{fig6}.  For
Fig.~\ref{fig5} a HTSE curve (calculated with $a_{n}$ values obtained
explicitly from summing the tabulation in \cite{daboul:04}) is also
included with the simulation data. This curve, calculated with the
known $13$ leading HTSE terms only, is essentially exact in the high
to moderate $\tau$ region. The numerical data are in excellent
agreement with the HTSE curve.  The fits to the ThL envelope data
correspond to
\begin{equation}
  \chi(\tau) = 0.94\tau^{-1.59}\left(1+0.0625\tau^{2.4}\right)
\end{equation}
and
\begin{equation}
  \xi(\tau) = 0.98\beta\tau^{-0.72}\left(1+0.017\tau^{2.4}\right)
\end{equation}
Thus the exponent estimates are $\gamma = 1.59(2)$ and $\nu = 0.72(1)$
so $\eta= 2-\gamma/\nu= -0.20(2)$. In Section~\ref{sec:XII} a detailed
discussion is given of the Gaussian HTSE estimates of
Ref.~\cite{daboul:04}.  For both $\gamma$ and $\nu$ the corrections to
scaling in the whole paramagnetic temperature region are weak. For
$\chi(\tau)$ the "effective" correction appears to be a sum of
high-order correction terms. Any correction with $\theta \approx 1$,
which might be expected from either the conformal correction or from a
leading analytic correction, seems to be negligible.

\begin{figure}
  \includegraphics[width=3.5in]{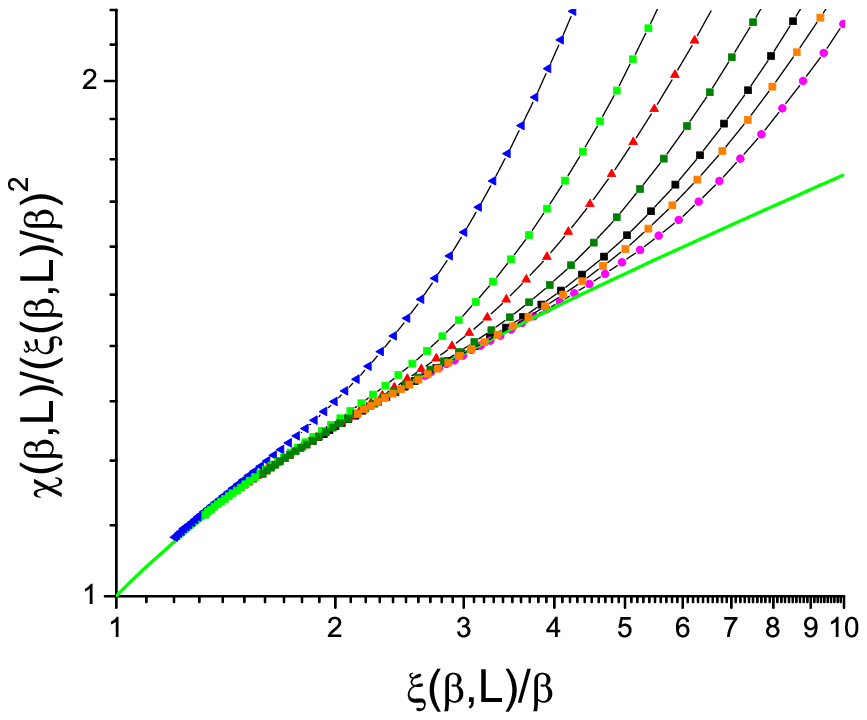}
  \caption{(Color on line) Gaussian 5D ISG. The ratio
    $\chi(\beta,L)/[\xi(\beta,L)/\beta]^2$ against
    $\xi(\beta,L)/\beta$ for $L = 10$, $9$, $8$, $7$, $6$, $5$ and $4$
    (right to left), continuous green curve : fit. No value is
    assumed for $\beta_{c}$. } \protect\label{fig7}
\end{figure}

A log-log plot of $y(\beta,L) =
\chi(\beta^2,L)/[\xi(\beta^2,L)/\beta]^2$ against $x(\beta,L) =
\xi(\beta^2,L)/\beta$ is shown in Fig.~\ref{fig7}. The estimated asymptotic
slope of the ThL envelope curve $\partial \ln y(\beta,L)/\partial
\ln x(\beta,L)$ gives an estimate for the critical exponent $\eta =
-0.19(2)$ without invoking any value for $\beta_{c}$. This $\eta$
estimate is consistent with the value from the ratio $\gamma/\nu$.

\begin{figure}
  \includegraphics[width=3.5in]{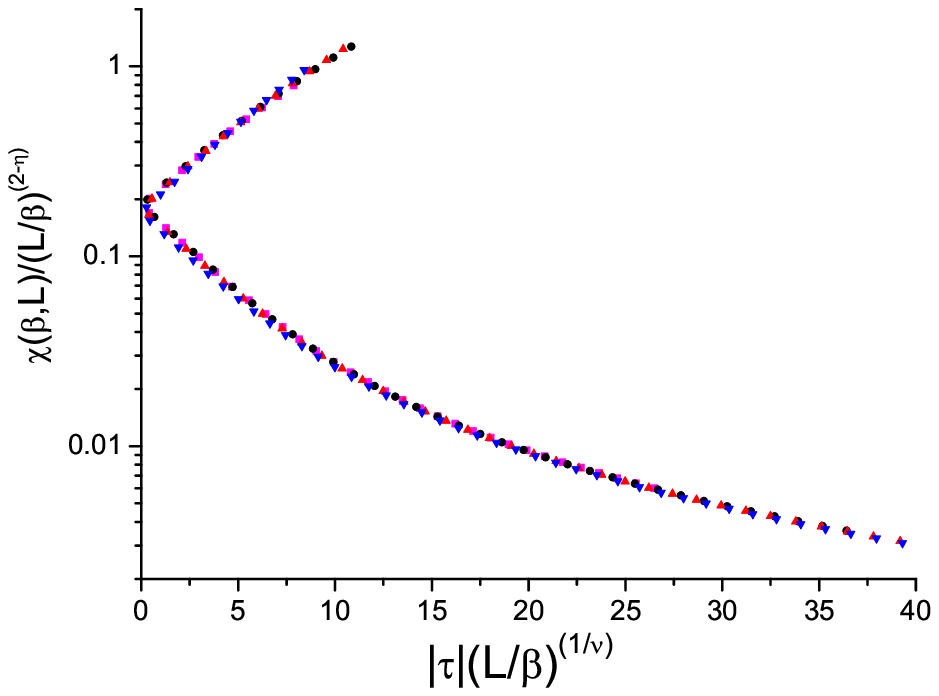}
  \caption{(Color on line) Gaussian 5D ISG. Privman-Fisher-like
    scaling of the $\chi(\beta,L)$ data following the form used in
    \cite{campbell:06}, with assumed parameters $\beta_{c}= 0.419$,
    $\nu = 0.72$, $\eta = -0.19$ and no adjustments.  $L=10$ pink
    squares, $L=8$ black circles, $L=6$ red triangles, $L=4$ blue
    inverted triangles.  Upper branch : $\beta > \beta_{c}$, lower
    branch $\beta < \beta_{c}$. } \protect\label{fig8}
\end{figure}

The basic Privman-Fisher extended scaling \eqref{PFextscal} for
$\chi(\beta,L)$ with these parameter values is shown in
Fig.~\ref{fig8}. The scaling is excellent (including the range of
temperatures below $T_{c}$, the upper branch) apart from weak
deviations visible for the smallest size $L=4$ which could be
accounted for by a finite size correction term.

\section{The 5D bimodal distribution ISG model}\label{sec:IX}
\begin{figure}
  \includegraphics[width=3.5in]{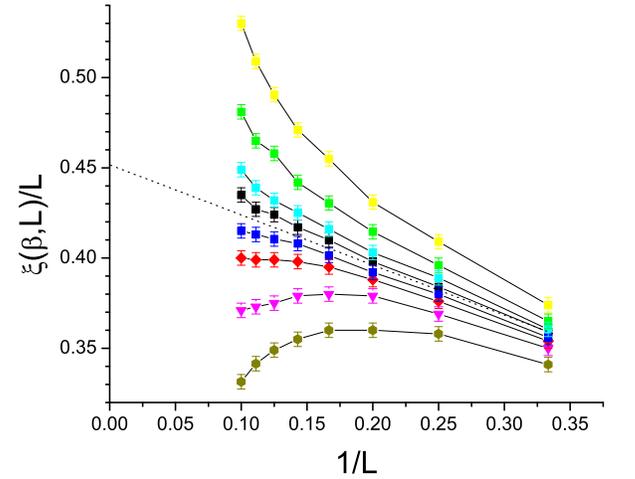}
  \caption{(Color on line) Bimodal 5D ISG. $\xi(\beta,L)/L$ against
    $1/L$ for fixed $\beta$, $\beta= 0.395$, $0.392$, $0.390$,
    $0.389$, $0.388$, $0.387$, $0.385$ and $0.382$ (top to bottom). $L
    = 10$, $9$, $8$, $7$, $6$, $5$, $4$ and $3$ (right to
    left). Dashed line : estimated criticality, $\beta = 0.3885$.}
  \protect\label{fig9}
\end{figure}

\begin{figure}
  \includegraphics[width=3.5in]{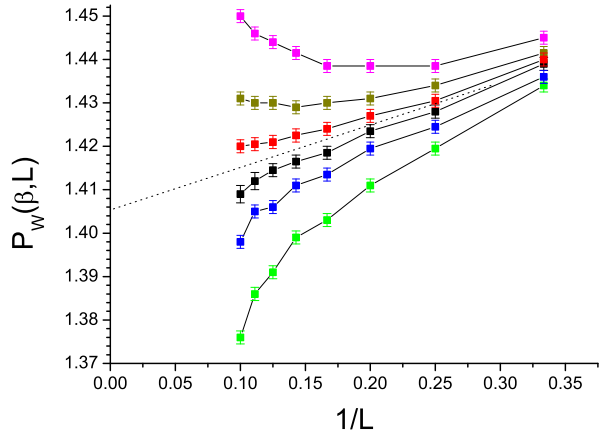}
  \caption{(Color on line) Bimodal 5D ISG. $P_{W}(\beta,L)$ against
    $1/L$ for fixed $\beta$, $\beta = 0.385$, $0.387$, $0.388$ ,
    $0.389$, $0.390$ and $0.392$ (top to bottom). $L = 10$, $9$, $8$,
    $7$, $6$, $5$, $4$ and $3$ (left to right).  Dashed line :
    estimated criticality, $\beta = 0.3885$.}  \protect\label{fig10}
\end{figure}

\begin{figure}
  \includegraphics[width=3.5in]{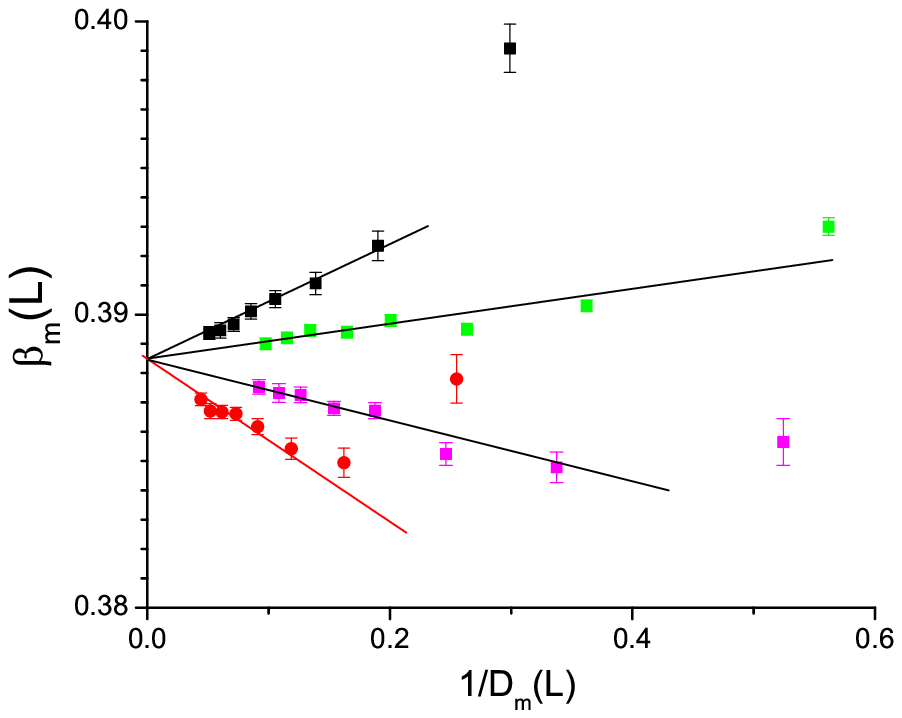}
  \caption{(Color on line) Bimodal 5D ISG.  Peak location $y =
    \beta_{m}$ against inverse peak height $x = 1/D_{m}$ for the
    derivative sets $\partial h(\beta,L)/\partial \beta$,
    $\partial P_{W}(\beta,L)/\partial \beta$, $\partial
    P_{\mathrm{skew}}(\beta,L)/\partial \beta$ and $\partial
    g(\beta,L)/\partial \beta$ (top to bottom).  Sizes $L= 3$, $4$,
    $5$, $6$, $7$, $8$, $9$ and $10$ (increasing to the left). For
    each observable the points extrapolate to $y(x) = \beta_{c}$ at
    the intercept, see text} \protect\label{fig11}
\end{figure}

For this model the dimensionless observable sets all show corrections
to finite size scaling. Data for two typical observables are shown in
Figs.~\ref{fig9} and \ref{fig10}. For $\beta_{c}$ the best overall
estimate is $\beta_{c}=0.3885(5)$.
Thermodynamic derivative peak location data are shown in Fig.~\ref{fig11}. The
extrapolations are consistent with the same value, $\beta_{c}=
0.3885(5)$.

\begin{figure}
  \includegraphics[width=3.5in]{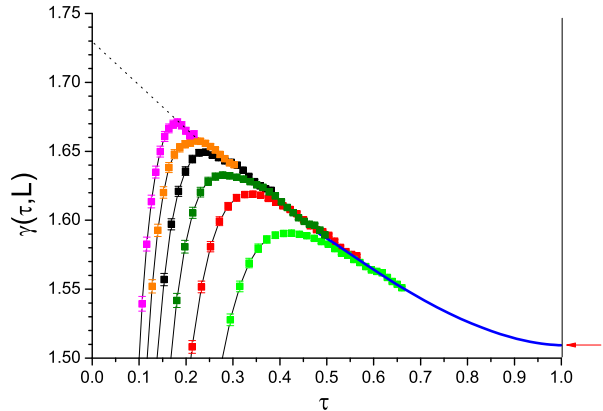}
  \caption{(Color on line) Bimodal 5D ISG. Effective exponent
    $\gamma(\tau,L)$ as function of $\tau$ with $\beta_{c} =
    0.3885$. Points : simulation data for $L=10$, $9$, $8$, $7$, $6$,
    $5$ (left to right), continuous (blue) curve on the right :
    calculated by summing the HTSE tabulation of
    \cite{daboul:04}. Dashed line : fit. } \protect\label{fig12}
\end{figure}

The effective exponents $\gamma(\tau,L)$ and $\nu(\tau,L)$ defined
above are shown in Fig.~\ref{fig12} and 13. The high temperature curve included
in Fig.~\ref{fig12} is evaluated from the HTSE series tabulation in
Ref.~\cite{daboul:04}. The critical exponents estimated by
extrapolation are $\gamma = 1.73(3)$ and $\nu = 0.76(1)$, and the fit
curves correspond to the ThL expressions
\begin{equation}
  \chi(\tau) = 0.73 \tau^{-1.73}\left(1+0.37\tau^{0.95}-0.005\tau^{8}\right)
\end{equation}
and
\begin{equation}
  \xi(\tau) = 0.94\beta\tau^{-0.76}\left(1+0.068\tau\right)
\end{equation}
The simulation $\beta_{c}$, $\gamma$ and $\nu$ values are in excellent
agreement with the quite independent HTSE bimodal critical value
estimates $\beta_{c}=0.389(1)$, $\gamma = 1.73(3)$, and $\nu \approx
0.73$ of Klein {\it et al} \cite{klein:91} discussed in detail in
Section~\ref{sec:XII}.

A log-log plot of $y(\beta,L) =
\chi(\beta^2,L)/[\xi(\beta^2,L)/\beta]^2$ against $x(\beta,L) =
\xi(\beta^2,L)/\beta$ is shown in Fig.~\ref{fig14}. The estimated limiting
slope of the ThL envelope curve $\partial \ln y(\beta,L)/\partial
\ln x(\beta,L)$ gives an estimate for the critical exponent $-\eta =
0.28(1)$ without invoking any estimate for $\beta_{c}$.
The Privman-Fisher extended scaling plot for $\chi(\beta,L)$ with
these critical parameters is shown in Fig.~\ref{fig15}.

\begin{figure}
  \includegraphics[width=3.5in]{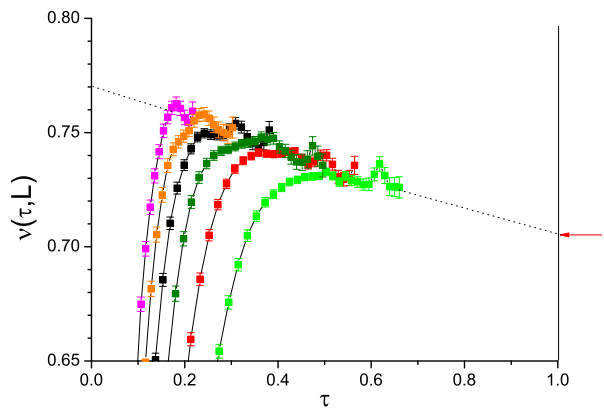}
  \caption{(Color on line) Bimodal $5d$ ISG. Effective exponent
    $\nu(\tau,L)$ as function of $\tau$ with $\beta_{c} =
    0.3885$. Points : simulation data for for $L=10$, $9$, $8$, $7$,
    $6$ and $5$ (left to right). Red arrow : exact limit.  Dashed line
    : fit}.  \protect\label{fig13}
\end{figure}

\begin{figure}
  \includegraphics[width=3.5in]{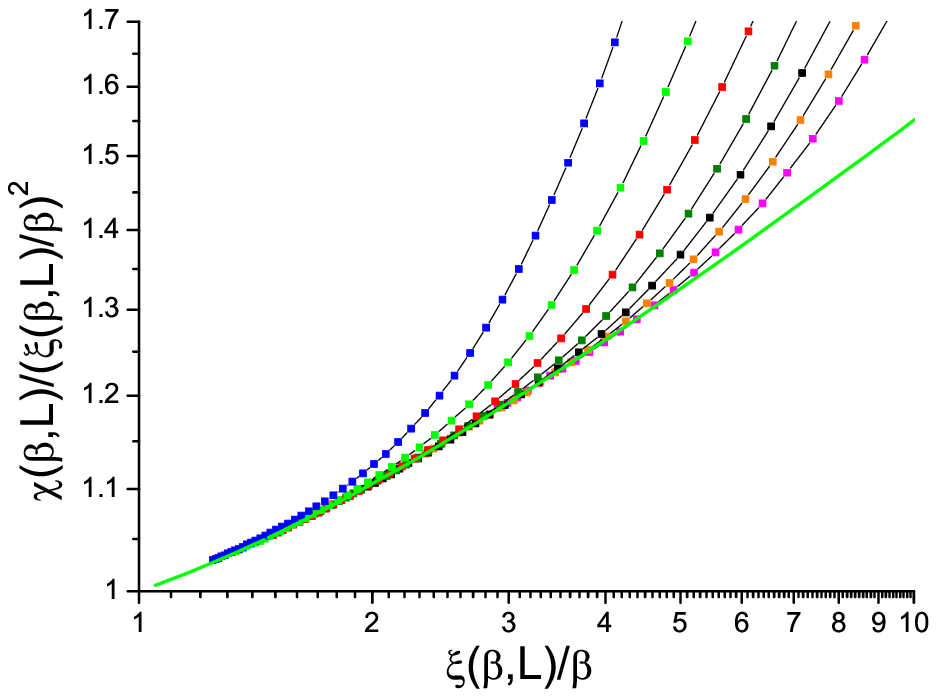}
  \caption{(Color on line) Bimodal 5D ISG. The ratio
    $\chi(\beta,L)/[\xi(\beta,L)/\beta]^2$ against
    $\xi(\beta,L)/\beta$ for $L = 10$, $9$, $8$, $7$, $6$, $5$ and $4$
    (right to left), continuous (green) curve : fit. No value is
    assumed for $\beta_{c}$.}  \protect\label{fig14}
\end{figure}

\begin{figure}
  \includegraphics[width=3.5in]{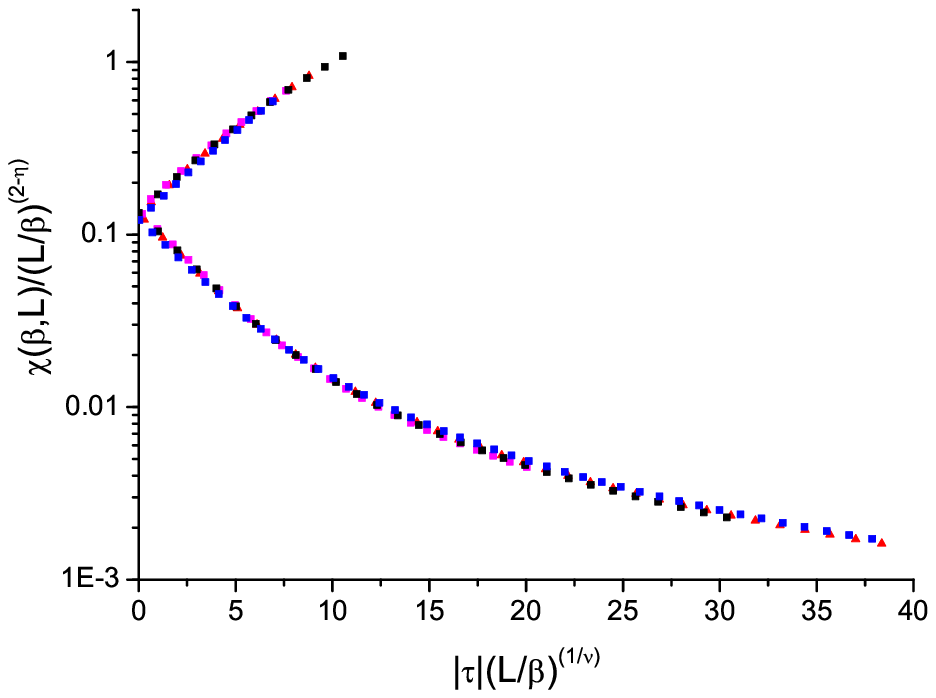}
  \caption{(Color on line) Bimodal $5d$ ISG. Privman-Fisher-like
    scaling of the $\chi(\beta,L)$ data following the form used in
    \cite{campbell:06}, with assumed parameters $\beta_{c}= 0.3885$,
    $\nu = 0.77$, $\eta = -0.25$ and no adjustments.  $L=10$ pink
    squares, $L=8$ black circles, $L=6$ red triangles, $L=4$ blue
    inverted triangles.  Upper branch : $\beta > \beta_{c}$, lower
    branch $\beta < \beta_{c}$.}  \protect\label{fig15}
\end{figure}

\section{The 5D uniform distribution ISG model}\label{sec:X}
The numerical data for the uniform distribution model and the
Laplacian distribution model are less complete than for the bimodal
and Gaussian models. Nevertheless reliable critical parameter
estimates have been obtained for both models.

\begin{figure}
  \includegraphics[width=3.5in]{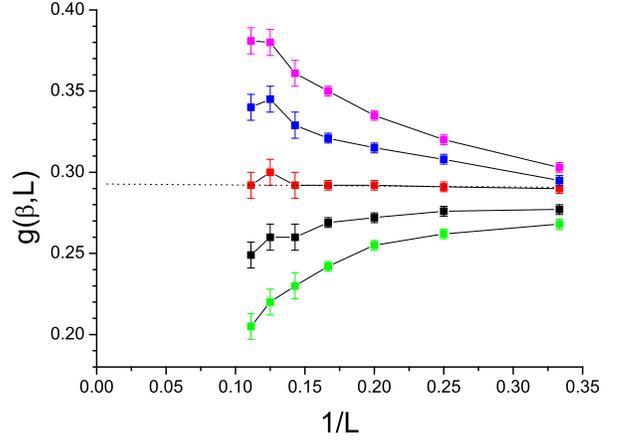}
  \caption{(Color on line) Uniform 5D ISG. The Binder cumulant
    $g(\beta,L)$ against $1/L$ for fixed $\beta$, $\beta = 0.405$,
    $0.4025$, $0.400$ , $0.3975$ and $0.395$ (top to bottom). $L = 9$,
    $8$, $7$, $6$, $5$, $4$ and $3$ (left to right).  Dashed line :
    estimated criticality, $\beta = 0.400$.}  \protect\label{fig16}
\end{figure}

\begin{figure}
  \includegraphics[width=3.5in]{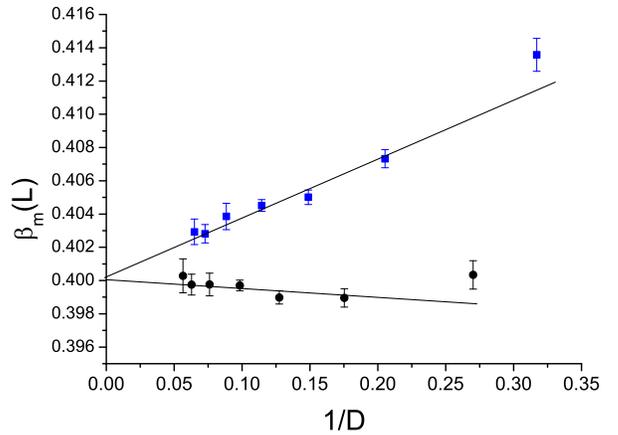}
  \caption{(Color on line) Uniform 5D ISG.  Peak location $y =
    \beta_{m}$ against inverse peak height $x = 1/D_{m}$ for the
    derivative sets $\partial h(\beta,L)/\partial \beta$ (top) and
    $\partial g(\beta,L)/\partial \beta$ (bottom).  Sizes $L= 3$, $4$,
    $5$, $6$, $7$, $8$ and $9$ (increasing to the left). For both
    observables the points extrapolate to $y(x) = \beta_{c}$ at the
    intercept, see text} \protect\label{fig17}
\end{figure}

\begin{figure}
  \includegraphics[width=3.5in]{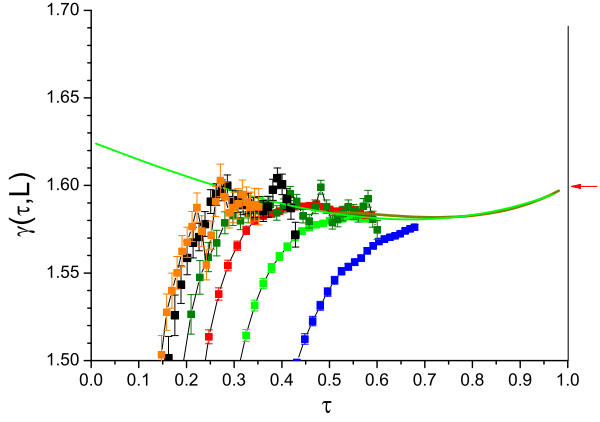}
  \caption{(Color on line) Uniform 5D ISG. Effective exponent
    $\gamma(\tau,L)$ as function of $\tau$ with $\beta_{c} =
    0.400$. Points : simulation data for $L = 9$, $8$, $7$, $6$, $5$
    and $4$ (left to right). Red arrow : exact limit.  Continuous
    (green) curve : fit. Continuous (red) curve on the right, almost
    hidden under the fit curve : calculated by summing the HTSE
    tabulation of \cite{daboul:04}. } \protect\label{fig18}
\end{figure}

\begin{figure}
  \includegraphics[width=3.5in]{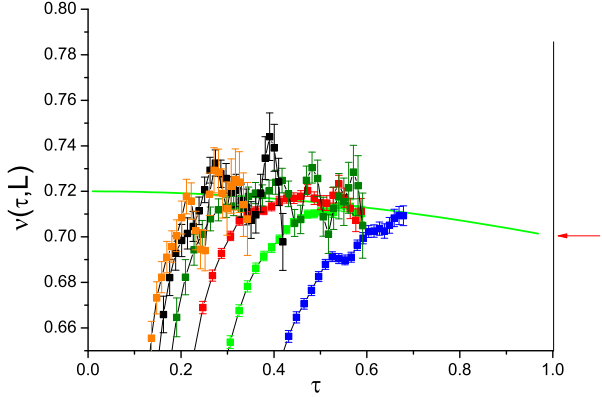}
  \caption{(Color on line) Uniform 5D ISG. Effective exponent
    $\nu(\tau,L)$ as function of $\tau$ with $\beta_{c} =
    0.400$. Points : simulation data for for $L= 9$, $8$, $7$, $6$ and
    $5$ (left to right). Red arrow : exact limit.  Continuous (green)
    curve : fit}.  \protect\label{fig19}
\end{figure}

\begin{figure}
  \includegraphics[width=3.5in]{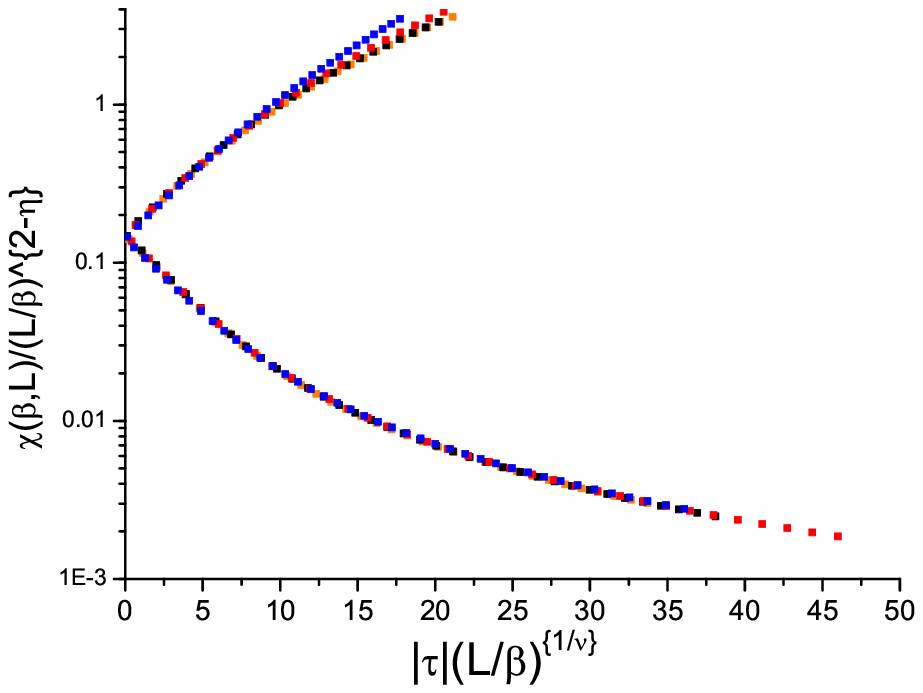}
  \caption{(Color on line) Uniform 5D ISG. Privman-Fisher-like
    scaling of the $\chi(\beta,L)$ data following the form used in
    \cite{campbell:06}, with assumed parameters $\beta_{c} = 0.3885$,
    $\nu = 0.77$, $\eta = -0.25$ and no adjustments.  $L=8$ black
    circles, $L=6$ red triangles, $L=4$ blue inverted triangles.
    Upper branch : $\beta > \beta_{c}$, lower branch $\beta <
    \beta_{c}$.}  \protect\label{fig20}
\end{figure}

For the uniform distribution model the FSS scaling data for the
dimensionless observables $P_{W}(\beta,L)$, $h(\beta,L)$ and
$W_{q}(\beta,L)$ all happen to show negligible corrections to scaling
and all consistently indicate $\beta_{c}= 0.400(1)$,
Fig.~\ref{fig16}. The data for the other dimensionless observables
show only weak corrections to scaling and are consistent with this
$\beta_{c}$.
The thermodynamic derivative peak data also confirm the critical
temperature value, Fig.~\ref{fig17}. The ThL effective exponent fits correspond
to
\begin{equation}
  \chi(t)=0.93\tau^{-1.625}\left(1+0.104 \tau-0.025 \tau^{3}\right)
\end{equation}
and
\begin{equation}
  \xi(\tau) = 0.99\tau^{-0.72}\left(1+0.01 \tau^{2.0}\right)
\end{equation}
so estimates $\gamma = 1.625(20)$, $\nu = 0.72(1)$ and $\eta =
-0.26(3)$, Figs.~\ref{fig18} and \ref{fig19}. The corrections to
scaling are weak. The $\beta_c$ and $\gamma$ values can be compared to
the HTSE estimates \cite{daboul:04} $\beta_{c} = 0.4016(37)$ and
$\gamma=1.70(15)$. (Here the critical temperature quoted is in terms
of the present normalization, not to that used in
Ref.~\cite{daboul:04}). The simulation and HTSE results are
consistent, with the wide error bar in the HTSE $\gamma$ being mainly
due to the associated uncertainty in the HTSE $\beta_{c}^2$.  The
Privman-Fisher extended scaling plot for $\chi(\beta,L)$ is shown in
Fig.~\ref{fig20}. The scaling is excellent until temperatures well
below $T_{c}$.

\section{The Laplacian distribution model}\label{sec:XI}
\begin{figure}
  \includegraphics[width=3.5in]{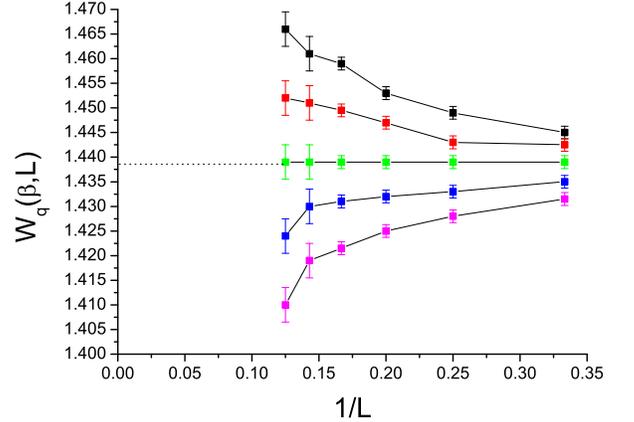}
  \caption{(Color on line) Laplacian 5D ISG. The parameter
    $W_{q}(\beta,L)$ against $1/L$ for fixed $\beta$, $\beta = 0.450$,
    $0.4525$, $0.455$, $0.4575$ and $0.460$ (top to bottom). $L = 8$,
    $7$, $6$, $5$, $4$ and $3$ (left to right).  Dashed line :
    estimated criticality, $\beta = 0.455$.}  \protect\label{fig21}
\end{figure}

\begin{figure}
  \includegraphics[width=3.5in]{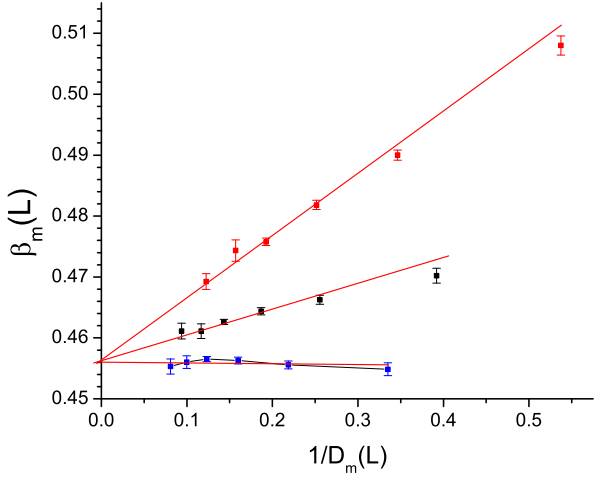}
  \caption{(Color on line) Laplacian 5D ISG.  Peak location $y =
    \beta_{m}$ against inverse peak height $x = 1/D_{m}$ for the
    derivative sets $\partial W_{q}(\beta,L)/\partial \beta$,
    $\partial h(\beta,L)/\partial \beta$ and $\partial
    g(\beta,L)/\partial \beta$ (top to bottom).  Sizes $L= 3$, $4$,
    $5$, $6$, $7$ and $8$ (increasing to the left). For each
    observables the points extrapolate to $y(x) = \beta_{c}$ at the
    intercept, see text} \protect\label{fig22}
\end{figure}

\begin{figure}
  \includegraphics[width=3.5in]{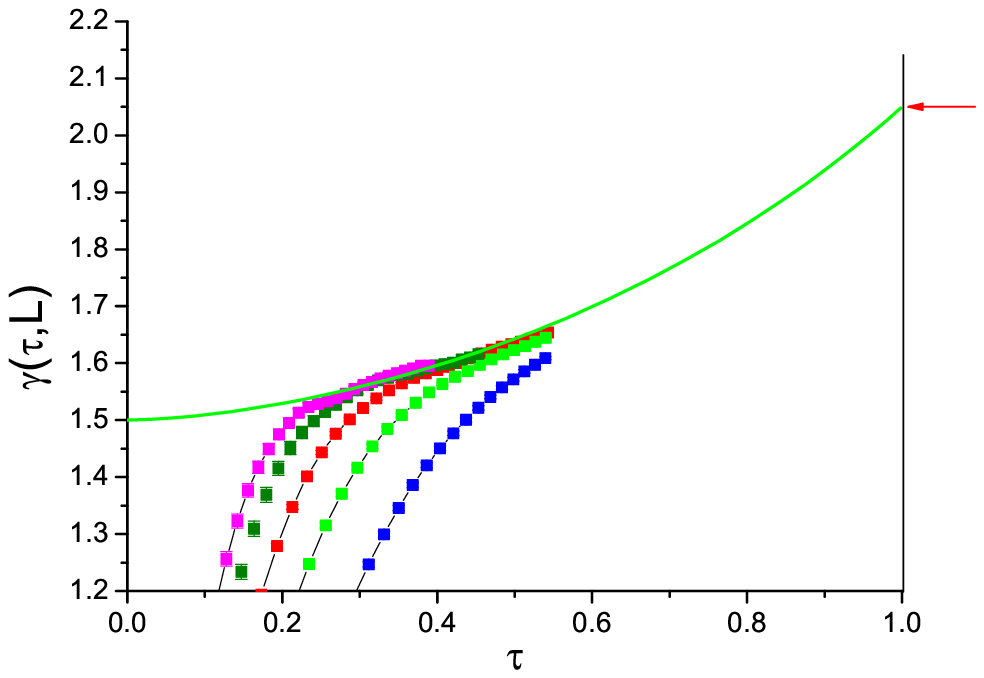}
  \caption{(Color on line) Laplacian 5D ISG. Effective exponent
    $\gamma(\tau,L)$ as function of $\tau$ with $\beta_{c} =
    0.455$. Points : simulation data for $L= 8$, $7$, $6$, $5$ and $4$
    (left to right). Red arrow : exact limit.  Continuous (green)
    curve : fit.  } \protect\label{fig23}
\end{figure}

For the Laplacian distribution model, the FSS $P_{W}(\beta,L)$ data
happen to show a negligible correction to scaling, Fig.~\ref{fig21},
providing an accurate estimate $\beta_{c}= 0.455(1)$. The data for the
other dimensionless observables show weak corrections to
scaling. Fixed temperature plots of the data, Figs.~\ref{fig22} and
\ref{fig23}, are consistent with the same $\beta_{c}$, and the
critical values of the dimensionless observables given in Table I.
The ThL data fits correspond to
\begin{equation}
  \chi(\tau) = 1.33\tau^{-1.5}\left(1-0.25\tau^{1.65}\right)
\end{equation}
and
\begin{equation}
  \xi(\tau) =0.973\beta\tau^{-0.69}\left(1+0.028\tau^{2.5}\right)
\end{equation}
leading to the critical parameter estimates $\gamma= 1.50(5)$, $\nu =
0.69(2)$ and $\eta = -0.17(3)$. The effective correction exponents are
relatively high indicating a low prefactor for a leading term with
$\theta \approx 1.0$.

\begin{figure}
  \includegraphics[width=3.5in]{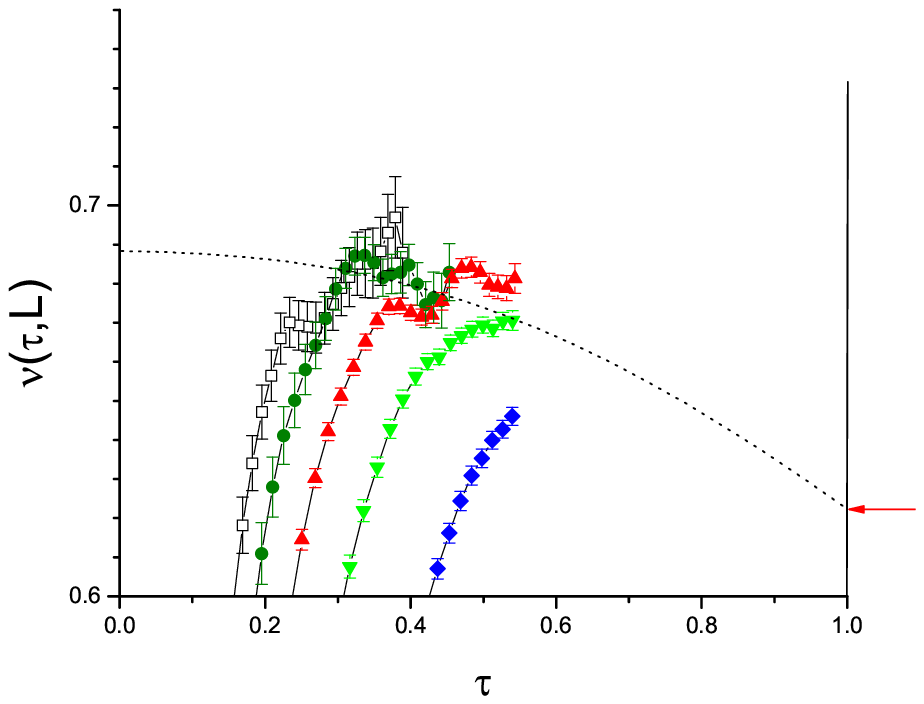}
  \caption{(Color on line) Laplacian 5D ISG. Effective exponent
    $\nu(\tau,L)$ as function of $\tau$ with $\beta_{c} =
    0.455$. Points : simulation data for for $L= 8$, $7$, $6$, $5$ and
    $4$ (left to right). Red arrow : exact limit.  Continuous (green)
    curve : fit}.  \protect\label{fig24}
\end{figure}

\begin{figure}
  \includegraphics[width=3.5in]{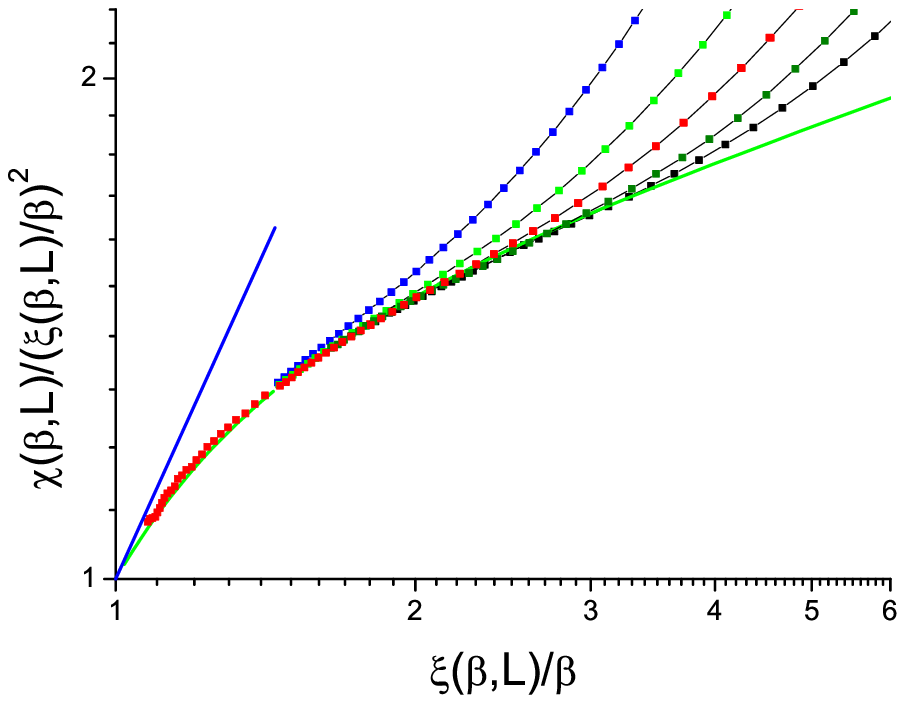}
  \caption{(Color on line) Laplacian 5D ISG. The ratio
    $\chi(\beta,L)/[\xi(\beta,L)/\beta]^2$ against
    $\xi(\beta,L)/\beta$, $L = 8$, $7$, $6$, $5$ and $4$ (right to
    left), continuous (green) curve : fit. No value is assumed for
    $\beta_{c}$.}  \protect\label{fig25}
\end{figure}

A log-log plot of $y(\beta,L) =
\chi(\beta^2,L)/[\xi(\beta^2,L)/\beta]^2$ against $x(\beta,L) =
\xi(\beta^2,L)/\beta$ is shown in Fig.~\ref{fig24}. The estimated
limiting slope of the ThL envelope curve $\partial \ln
y(\beta,L)/\partial \ln x(\beta,L)$ gives an estimate for the critical
exponent $\eta = -0.19(3)$ without invoking any estimate for
$\beta_{c}$.  The Privman-Fisher extended scaling for $\chi(\beta,L)$
is shown in Fig.~\ref{fig25}.  There are no published HTSE data on
this model.

\begin{figure}
  \includegraphics[width=3.5in]{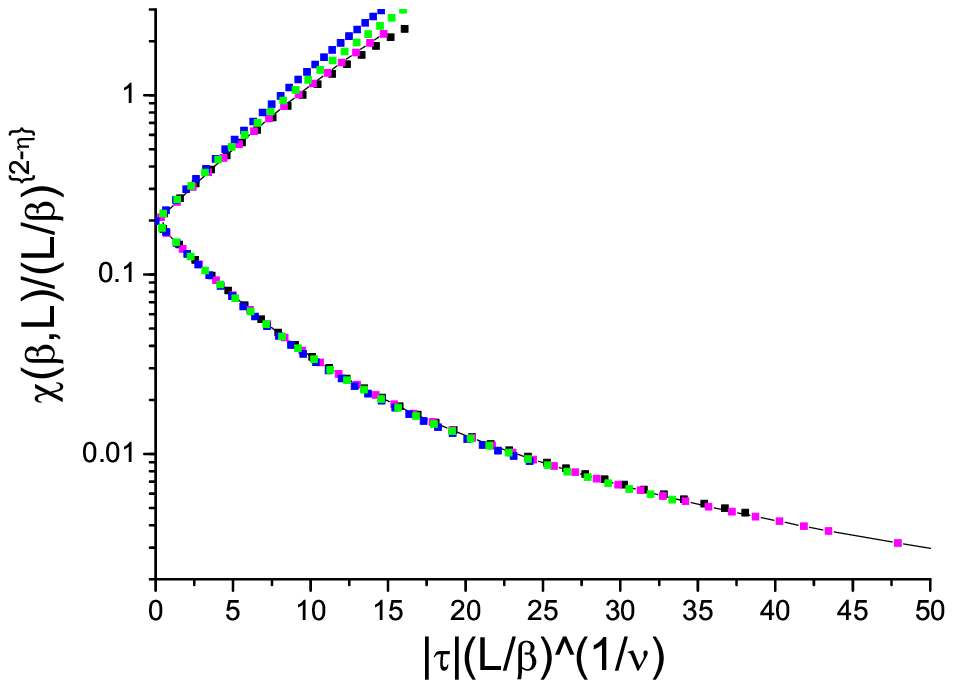}
  \caption{(Color on line) Laplacian 5D ISG. Privman-Fisher-like
    scaling of the $\chi(\beta,L)$ data following the form used in
    \cite{campbell:06}, with assumed parameters $\beta_{c}= 0.455$,
    $\nu = 0.69$, $\eta = -0.21$ and no adjustments.  $L=8$ black
    squares, $L=6$ red circles, $L=5$ green
    triangles, $L=4$ blue inverted triangles.  Upper branch : $\beta >
    \beta_{c}$, lower branch $\beta < \beta_{c}$.}
  \protect\label{fig26}
\end{figure}

\section{High Temperature Series Expansions}\label{sec:XII}
Having the numerical analyses in hand we will now discuss in detail
the HTSE data \cite{klein:91,daboul:04} published some years ago. The
HTSE technique is efficient for ISGs in dimension $5$ because of the
proximity to the ISG upper critical dimension $d = 6$.  High
temperature series expansion calculations have been made on the
bimodal ISG \cite{klein:91} in general dimension, using
$w=\tanh(\beta)^2$ as the scaling variable, and on ISGs with bimodal,
Gaussian, uniform and double triangle distributions using $\beta^2$ as
the scaling variable \cite{daboul:04}, again in general dimension. The
number of series terms $a_{n}$ evaluated was limited by practical
considerations to $n=15$ for bimodal interactions in both cases and to
$n=13$ for the other distributions \cite{daboul:04}.

In Ref.~\cite{daboul:04} the spin-glass susceptibility terms were
evaluated, and the series were analyzed through Dlog Pad\'e, $M1$ and
$M2$ techniques combined with Euler-transformations (see
Ref.~\cite{daboul:04} for details concerning these techniques). The
precision on the extrapolations to criticality was limited by the
restricted number of terms, and by a parasitic antiferromagnetic
contribution which oscillates in sign and grows in strength with
increasing $n$. (The Euler transformation is designed to reduce the
influence of this parasitic term). The critical $\beta_{c}^2$, the
critical exponent $\gamma$, and the leading correction term exponent
$\theta$ were evaluated globally using the different analysis
techniques. The final estimates for both $\beta_{c}^2$ and $\gamma$
were cited with rather large error bars.  We will concentrate on the
Dlog Pad\'e analysis. Including Euler transformations, a large number
of individual Dlog Pad\'e solutions were generated for each
model. Each individual solution provided precise linked estimates of
the critical parameters $[\beta_{c}^2, \gamma]$.  For the 5D Gaussian
model explicit point by point data were presented in Fig.~7 of
Ref.~\cite{daboul:04}, which shows the $\gamma$ against $\beta_{c}^2$
estimates for each individual solution. The values of the two
parameters are highly correlated, with the estimates being fairly
dispersed, but with the $\gamma$ values essentially a smooth function
of the $\beta_{c}^2$ values (see inset to Fig.~7 of
Ref.~\cite{daboul:04}). The authors quote as their final Dlog Pad\'e
estimates $\beta_{c}^2 \approx 0.174$ with the associated global
estimate $\gamma = 1.67(8)$, and $\beta_{c}^2 = 0.177(3)$ and
$1.75(15)$ from the other analyses, together with $\theta \approx 1.0$
from all techniques.  Imposing the present accurate simulation
estimate $\beta_{c}^2 = 0.1755(5)$ from FSS and thermodynamic
derivative peak analyses onto the Gaussian Dlog-Pad\'e results in the
inset to Fig.~7 of \cite{daboul:04}, one can read off a
corresponding "threshold biased" estimate $\gamma = 1.59(2)$. This is
in full agreement with the Gaussian model simulation estimate above,
$\gamma = 1.60(1)$. Unfortunately no point by point Dlog Pad\'e
figures equivalent to that for the Gaussian model were presented for
the bimodal model or for the uniform model.

For the bimodal model in dimension $5$, the HTSE estimates in
\cite{daboul:04} are $\beta_{c}^2 = 0.154(3)$, $\gamma =1.91(10)$ or
$1.95(15)$, again with rather wide error bars.  However the earlier
HTSE study by the same group on the bimodal ISG model in general
dimension \cite{klein:91} using $w = \tanh(\beta)^2$ as scaling
parameter was more complete than that of \cite{daboul:04}, because in
addition to the series for the spin-glass susceptibility (referred to
as $\Gamma_{2}$ in Ref.~\cite{klein:91}), series for the two higher
order susceptibilities $\Gamma_{3}$ and $\Gamma_{4}$ (defined in
\cite{klein:91}) were also evaluated. The RGT critical exponents for
these higher order susceptibilities are $\gamma_{3} =
(3\gamma+d\nu)/2$ and $\gamma_{4} = 2\gamma + d\nu$. We have evaluated
explicitly the terms $a_{n}$ for the different series from the
tabulations given in Ref.~\cite{klein:91}. It turns out that in
dimension $5$ the parasitic oscillating terms in the $a_{n}$ series
are much weaker for these higher order susceptibilities than for the
standard ISG susceptibility.  Because of the supplementary information
from the higher order susceptibilities, the estimates for the critical
temperature and the critical exponents in the dimension $5$ bimodal
ISG model are much more precise in Ref.~\cite{klein:91} than in
\cite{daboul:04}.  The final estimates presented in
Ref.~\cite{klein:91} are $w_{c} = 0.1372(8)$, i.e. $\beta_{c} =
0.389(1)$ or $\beta_{c}^2 = 0.1513(8)$, and $\gamma = 1.73(3)$,
$\gamma_{3} = 4.4(1)$, and $\gamma_{4} = 7.3(2)$ together with $\theta
\approx 1.0$. These values can be compared with the independent values
from the simulation estimates given above : $\beta_{c} = 0.3885(5)$,
$\gamma = 1.73(2)$, $\gamma_{3} = (3\gamma+d\nu)/2 = 4.5(1)$,
$\gamma_{4} = 2\gamma + d\nu = 7.3(2)$ and $\theta \approx 1.0$.
Remarkably, the present 5D bimodal estimates, based on data obtained
from the simulation approach which is entirely independent technically
from HTSE, are in uncanny agreement with the HTSE estimates from $25$
years ago.

For the 5D uniform model the estimates in Ref.~\cite{daboul:04} are
$\beta_{c}^2 = 0.162(3)$ (with the present normalization) and $\gamma
= 1.70(15)$, compatible with but less accurate than the the simulation
estimates $\beta_{c}^2 = 0.160(1)$ and $\gamma=1.66(2)$. A threshold
biased HTSE Dlog Pad\'e estimate for $\gamma$ would certainly reduce
the wide error bar if individual Dlog Pad\'e estimates were available.
No HTSE studies have been made of the 5D Laplacian model.

It is important that both Ref.~\cite{daboul:04} and \cite{klein:91}
estimate the correction exponent in dimension $5$ to be $\theta
\approx 1.0$ for all models. By definition there can be correction
terms with higher exponents but no correction term with a lower
exponent. The corresponding finite size correction exponent estimate
is $\omega = \theta/\nu \approx 1.2$.  These HTSE bimodal and
threshold biased Gaussian $\gamma$ estimates ($1.73(3)$ and $1.60(2)$
respectively) confirm the non-universality of 5D ISG critical
exponents.

\section{Conclusion}\label{sec:XIII}
The critical temperatures, critical exponents, and critical values for
a number of dimensionless observables, have been estimated for the
bimodal, Gaussian, uniform and Laplacian distribution ISG models in
dimension $5$ from numerical simulations. The values are summarized in
Table I.

\begin{table}[htbp]
  \caption{\label{Table:I} Estimates of the critical inverse
    temperatures, exponents and critical dimensionless parameters
    $\beta_{c}$, $\gamma$, $\nu$, $\eta$, $g(\beta_{c})$,
    $\xi/L(\beta,L)$, $h(\beta,L)$, $W_{q}(\beta,L)$ and for $5$D
    bimodal, uniform, Gaussian and Laplacian distribution ISG models.}
  \begin{ruledtabular}
    \begin{tabular}{ccccc}
    model&bimodal&uniform&Gaussian&Laplacian \\
    Kurtosis&$1$&$1.8$&$3$&$6$ \\
    $\beta_{c}$&$0.3885(5)$&$0.4000(5)$&$0.4190(5)$&$0.455(1)$ \\
    $\gamma$&$1.73(2)$&$1.625(20)$&$1.600(5)$&$1.49(2)$ \\
    $\nu$&$0.77(2)$&$0.72(1)$&$0.720(5)$&$0.69(1)$ \\
    $\eta$&$-0.25(3)$&$-0.26(3)$&$-0.22(2)$&$-0.21(2)$ \\
    $g(\beta_c)$&$0.34(1)$&$0.29(1)$&$0.300(5)$&$0.265(5)$ \\
    $\xi/L(\beta_c)$&$0.450(5)$&$0.42(1)$&$0.425(3)$&$0.401(3)$\\
    $P_{W}(\beta_c)$&$1.415(10)$&$1.425(5)$ &$1.425(10)$&$1.438(4)$ \\
    $P_{\mathrm{skew}}(\beta_c)$&$1.41(1)$&$1.422(10)$ &$1.422(10)$&$1.442(2)$ \\
    $W_{q}(\beta_c)$&$0.155(5)$&$0.125(2)$ &$0.128(2)$&$0.115(3)$ \\
    $h(\beta_c)$&$0.260(5)$&$0.225(2)$ &$0.230(2)$&$0.215(2)$ \\
    \end{tabular}
  \end{ruledtabular}
\end{table}


The accurate ISG inverse ordering temperature $\beta_{c}$ values in
5D increase regularly with the kurtosis $K$ of the interaction
distribution, in agreement with earlier HTSE estimates and as expected
from basic physical arguments \cite{singh:88,campbell:05}.

More remarkably, the critical exponents also evolve regularly with
$K$.  As $K$ increases, the critical exponents $\gamma$ and $\nu$
decrease regularly. Thus the uniform, Gaussian and Laplacian model
$\gamma$ estimates are approximately $4\%$, $8\%$ and $15\%$
respectively below the bimodal value. The critical values of the
dimensionless parameters also vary if not quite so regularly; the
critical dimensionless observable values for the extreme models
(bimodal and Laplacian) differ by up to about $30\%$ depending on the
observable.

Comparisons are made between the present simulation estimates for the
exponent $\gamma$ in the bimodal and Gaussian models, and those
obtained independently from HTSE. The most accurate published HTSE
bimodal model $\beta_{c}$ and $\gamma$ values \cite{klein:91} and the
present simulation estimates are in full agreement, $\beta_{c}=
0.3885(5)$ and $\gamma = 1.73(3)$. In the Gaussian model, if the
present precise simulation value for $\beta_{c}$ is used to threshold
bias the analysis of the HTSE data \cite{daboul:04}, the HTSE $\gamma$
value fully agrees with the simulation estimate. Both techniques then
give as the Gaussian model estimate $\gamma = 1.60(2)$, so clearly
lower than the bimodal model value.

These dimension $d=5$ ISG data thus confirm the empirical conclusion
reached from dimension $d=4$ and dimension $d=2$ studies
\cite{lundow:15,lundow:15a,lundow:16} that ISG models in a fixed
dimension but with different interaction distributions do not lie in
the same universality class.

It is relevant that experimental measurements have already shown
clearly that critical exponents in $d=3$ Heisenberg spin glasses vary
considerably from system to system, depending on the strength of the
Dzyaloshinsky-Moriya coupling term \cite{campbell:10}.


\begin{acknowledgments}
  The authors wish to thank Ralph Chamberlin, Joes Bijvoet and Jan
  Aarts for helpful comments.  The computations were performed on
  resources provided by the Swedish National Infrastructure for
  Computing (SNIC) at the High Performance Computing Center North
  (HPC2N) and Chalmers Centre for Computational Science and
  Engineering (C3SE).
\end{acknowledgments}

\end{document}